\documentclass[prd,onecolumn,showpacs,preprintnumbers,amsmath,amssymb,floatfix,nofootinbib]{revtex4}

\usepackage{graphicx}
\usepackage{dcolumn}
\usepackage{xspace}
\usepackage[utf8]{inputenc}
\usepackage{float}
\usepackage{subfigure}
\usepackage{url}

\usepackage{color}
\usepackage{comment}
\usepackage{bbm}

\definecolor{BlueViolet}{rgb}{0.2, 0.00, 0.7}
\definecolor{Blue}{rgb}{0.15, 0.00, 0.9}
\definecolor{halayaube}{rgb}{0.4, 0.22, 0.33}
\definecolor{sanddune}{rgb}{0.59, 0.44, 0.09}
\usepackage[colorlinks=true,linkcolor=Blue,citecolor=Blue,urlcolor=BlueViolet,
hyperfootnotes=false]{hyperref}

\begin{document}

\preprint{}
\preprint{}
\title{Lepton flavor universality violation in semileptonic tree level weak transitions}

\author{K.~Azizi}
\email{ kazizi@dogus.edu.tr}
\affiliation{Physics Department, Do\u gu\c s University,
Ac{\i}badem-Kad{\i}k\"oy, 34722 Istanbul, Turkey}
\affiliation{Department of Physics, University of Tehran, North Karegar Ave., Tehran 14395-547, Iran}

\author{Y.~Sarac}
\email{yasemin.sarac@atilim.edu.tr}
\affiliation{Electrical and Electronics Engineering Department,
Atilim University, 06836 Ankara, Turkey}
\author{H.~Sundu}
\email{ hayriye.sundu@kocaeli.edu.tr}
\affiliation{Department of Physics, Kocaeli University, 41380 Izmit, Turkey}

\date{\today}

\begin{abstract}
The recent deviations of the experimental data on some parameters of the tree-level semileptonic  $B$ and $B_c$ mesons decays from the standard model (SM) predictions indicate considerable violations of the lepton flavor universality, and as a result possible new physics (NP) effects. To better understand  the possible NP effects it is necessary to study deeply the physical quantities defining these decays from many aspects. The calculations of the physical quantities require the determinations of the hadronic form factors entering the matrix elements of the considered transitions as the main inputs. We calculate the form factors governing the tree-level $B_c\rightarrow J/\psi l \nu$ and $B_c \rightarrow \eta_c l \nu$  transitions within the QCD sum rules method. The obtained form factors are  used in the calculations of the branching ratios ($BR$s) of the $B_c \rightarrow J/\psi l \nu$ and  $B_c \rightarrow \eta_cl \nu$  transitions as well as  $R(J/\psi)$ and $R(\eta_c)$. Our result on $R(J/\psi)$ supports the present tension between the SM theory prediction and the experimental data. Our result on $R(\eta_c)$ can be checked in future experiments.  
\end{abstract}
\maketitle

\vspace{-1mm}
\maketitle
\renewcommand{\thefootnote}{\#\arabic{footnote}}
\setcounter{footnote}{0}
\section{\label{sec:level1}Introduction}

Although the SM provides us with many predictions consistent with the experimental observations, there exist experimental and theoretical reasons to believe that it is not the ultimate theory of nature, but an effective theory. There are many issues that cannot be addressed by the SM. Motivated by this, some new models containing new particles or new interactions are proposed trying to find answers to these problems. The signatures of these particles are simultaneously investigated in experiments. Beside the direct searches at colliders, as an indirect approach for the investigation of new physics effects, the semi-leptonic decays involving $b\rightarrow c$ and $b\rightarrow s$ transitions provide crucial testing ground. Experimental results presented by the BABAR, Belle and LHCb collaborations~\cite{Lees:2012xj,Lees:2013uzd,Huschle:2015rga,Hirose:2016wfn,Hirose:2017dxl,Aaij:2015yra,Aaij:2017uff,Aaij:2017deq} have indicated serious deviations from the predictions of SM and triggered the interest on these types of decays. Naturally, the different masses of the charged leptons lead to differences in the branching ratios  of the decays containing these particles. However, further deviations from predictions of the SM imply the lepton flavor universality violation (LFUV) and make the subject intriguing from the point of NP effects investigations. Because of the higher mass of the involved $\tau$ lepton, the semileptonic transitions containing $\tau$ lepton have more sensitivity to the NP effects compared to the other leptons. As a result, a deeper understanding of these transitions will be helpful to test the SM and physics beyond it and improve our knowledge about its parameters.   

In this respect, investigations of the ratios of the branching fractions for the tree-level semileptonic transitions $B \rightarrow D ^{(*)}\tau  \nu$ to $B \rightarrow D^{(*)}  l  \nu$ or  $B_{c}\rightarrow J/\psi (\eta_c)\tau  \nu$ to $B_{c} \rightarrow J/\psi (\eta_c) l  \nu$, where $l$ is $\mu$ or $e$, will be helpful due to the reduction of the uncertainties coming from the hadronic transition form factors and cancellation of the Cabibbo-Kobayashi-Maskawa matrix elements. Our focus in the present study will be on $B_c \rightarrow J/\psi l \nu$ and  $B_c \rightarrow \eta_cl \nu$  transitions as well as  $R(J/\psi)$ and $R(\eta_c)$, however, in order to compare the order of experimental/theoretical uncertainties in $B$ and $B_c$ decays as well as the theory-experiment tensions, we  give the average results on $R(D)$ and $R(D^*)$ in the following, as well.

The experimental searches on $R(D)$ and $R(D^*)$  have leaded to the results with global average as~\cite{HFLAV:average} 
\begin{eqnarray}
R(D)=\frac{BR(B\rightarrow D \tau \nu_{\tau})}{BR(B\rightarrow D l \nu_{l})}=0.407\pm0.039\pm0.024 ,
\end{eqnarray}
and 
\begin{eqnarray}
R(D^*)=\frac{BR(B\rightarrow D^* \tau \nu_{\tau})}{BR(B\rightarrow D^* l \nu_{l})}=0.306\pm0.013\pm0.007,
\end{eqnarray}
while the existing predictions of the same ratios in SM are~\cite{HFLAV:average}
\begin{eqnarray}
R(D)=\frac{BR(B\rightarrow D \tau \nu_{\tau})}{BR(B\rightarrow D l \nu_{l})}=0.299\pm0.003 ,
\end{eqnarray}
and 
\begin{eqnarray}
R(D^*)=\frac{BR(B\rightarrow D^* \tau \nu_{\tau})}{BR(B\rightarrow D^* l \nu_{l})}=0.258\pm0.005 ,
\end{eqnarray}
indicating  deviations from the  experimental observations  at $3.8 \sigma$ level. 

%
%
%
%
%
%

 Coming back to $B_c \rightarrow J/\psi l \nu$ and  $B_c \rightarrow \eta_cl \nu$  channels, the LHCb has measured $ R(J/\psi) $ as~\cite{Aaij:2017tyk}
\begin{eqnarray}
R(J/\psi)=\frac{BR(B_c\rightarrow J/\psi \tau \nu_{\tau})}{BR(B_c\rightarrow J/\psi \mu \nu_{\mu})}=0.71 \pm 0.17 \pm 0.18,
\end{eqnarray}
having up to $\sim 2 \sigma$ deviations from the values predicted by the SM \cite{Cohen:2018dgz,Wen-Fei:2013uea,Rui:2016opu,Dutta:2017xmj,Leljak:2019eyw,Murphy:2018sqg,Issadykov:2018myx,Huang:2007kb,Kiselev:2002vz,Ivanov:2006ni,Wang:2008xt,Hu:2019qcn,Huang:2018nnq,Alok:2017qsi}. As it is seen, in the case of the tree-level $ b\rightarrow c $ transitions, the theory-experiment tension seems to be more serious in $ B $ meson decay channels compared to  those of the $ B_c $ meson.
As is also seen,  the experimental result on $ R(J/\psi) $ contains large errors compared to the ones in $ R(D) $ and $ R(D^*) $. Existing theoretical predictions on $ R(J/\psi) $ include larger uncertainties compared to $ R(D) $ and $ R(D^*) $, as well. The prediction of Ref. \cite{Cohen:2018dgz} as one of the recent and complete estimations  on $ R(J/\psi) $, i.e.,
\begin{eqnarray}
0.20\leqslant R(J/\psi)\leqslant 0.39,
\label{rjpsiteori}
\end{eqnarray}
 indicates a wide band to the value of this parameter. The model-independent bound in this study was constructed  by constraining the form factors through a combination of dispersive relations, heavy-quark relations at zero-recoil, and the limited existing determinations from lattice QCD. Thus, more precise theoretical predictions on $ R(J/\psi) $ and the related form factors are needed.  
Though measuring the similar ratio $R(\eta_c)$ is more difficult compared to $J/\psi$ case, this ratio and the related decay channels may be studied in near future, and therefore, providing detailed theoretical investigations will be helpful to gain deeper understanding about it and may shed light on the corresponding experiments. 

 The matrix elements of the semileptonic decays of the $B_c$ meson to $J/\psi l \nu$ and $\eta_c l \nu$ final states can be factorized to the leptonic and hadronic parts and, for the theoretical analysis, it is essential to know the corresponding hadronic transition form factors. Therefore, we focus on the matrix elements representing these hadronic transitions and calculate  the corresponding form factors. In literature, one can find various calculations on some of these form factors which were obtained via different methods. Some of these methods are the light cone QCD sum rules~\cite{Leljak:2019eyw,Fu:2018vap}, QCD sum rules~\cite{Kiselev:1999sc,Kiselev:2002vz,Colangelo:1}, Bethe-Salpeter equation~\cite{AbdElHady:1999xh}, perturbative QCD factorization approach~\cite{Wen-Fei:2013uea,Hu:2019qcn}, nonrelativistic QCD approach~\cite{Zhu:2017lqu}, covariant light-front quark model~\cite{Wang:2008xt}, covariant confined quark model~\cite{Tran:2018kuv,Issadykov:2018myx}, relativistic quark model~\cite{Ebert:2003cn} and nonrelativistic constituent quark model~\cite{Hernandez:2006gt}. To achieve the form factors of the related transitions in full theory, we employ the three point QCD sum rule~\cite{Ek1,Ek2,Ek3}, which is a powerful nonperturbative method applied in many calculations, successfully. The obtained form factors are used in the calculations of the decay widths and branching ratios of the considered decays as well as $ R(J/\psi) $ and $R(\eta_c)$. Our prediction on $ R(J/\psi) $ is compared with the present experimental data as well as the existing theoretical predictions. We also compare our results on the branching fractions of the $B_c \rightarrow J/\psi l \nu$ transitions with the existing theoretical estimations. Detailed information on the form factors and the ratio of $BR$s corresponding to the transition of $B_c$ to $\eta_c l \nu$ may also provide valuable insights for the future observations related to this channel and contribute to the investigations of NP effects. We compare our predictions on the branching fractions of the $B_c \rightarrow \eta_cl \nu$ transitions as well as $R(\eta_c)$ with the existing theoretical predictions. Note that, beside the calculations of $ R(J/\psi) $
and $R(\eta_c)$, which  contain small errors due to some cancellations, we calculate the individual branching ratios at each channel, as well. With the resent progresses in the experimental side we hope that we will be able to measure these branching fractions in near future. Comparison of the future data on the individual $BR$s with the results of the present study can help us constrain the SM parameters entering the calculations and get useful information about the form factors representing the decays under consideration.

The outline of the paper is as follows: In Section II, we provide the details of the calculations for the form factors of $B_c \rightarrow J/\psi l \nu$ and  $B_c \rightarrow \eta_cl \nu$ transitions in full theory. Section III is devoted to numerical analysis of these form factors and the calculations of the $BR$s of the considered decay channels. In this section, we also provide ratios, $R(J/\psi)$  and $R(\eta_c)$. The last section presents comparison of the results with existing theoretical and experimental information as well as our concluding remarks.

\section{Form Factors of $B_c\rightarrow\ J/\psi l \nu$ and $B_c\rightarrow \eta_c l \nu $ transitions \label{sec:QCDsumrules}}

In this section the form factors corresponding to the tree-level $B_c\rightarrow\ J/\psi l \nu$ and $B_c\rightarrow \eta_c l \nu $ transitions are calculated via three point QCD sum rules. For the considered transitions, the three point correlation function is as follows
\begin{eqnarray}
\Pi_{\mu\nu(\nu)}=i^2\int d^4x e^{-ipx}\int d^4y e^{ip'y}\langle0|\mathcal{T}\{ J_{J/ \psi,\mu}(J_{\eta_c})(y)J_{\nu}^{tr}(0)J^{\dagger}_{B_c}(x)\}|0\rangle,\label{Correlator}
\end{eqnarray}
where $J_{\nu}^{tr}(0)=\overline{c}(0)\gamma_{\nu}(1-\gamma_5) b(0)$ is the transition current and the interpolating currents of the participating mesons are given as
\begin{eqnarray}
J_{B_c}(x)=\overline{c}(x)\gamma_5 b(x),\nonumber\\
J_{J/ \psi,\mu}(y)=\overline{c}(y)\gamma_{\mu} c(y),\nonumber\\
J_{\eta_c}(y)=\overline{c}(y)\gamma_5 c(y).\label{Currents}
\end{eqnarray}
For the calculation of the correlation function two ways, whose results are matched at the end, are followed. Firstly, it is calculated in terms of the hadronic degrees of freedom such as the masses, decay constants and form factors. In this part of the calculations, complete sets of hadronic states carrying the same quantum numbers as the considered hadrons are inserted into the correlation function. This is followed by the integration and isolation of the ground state contributions, which turns the correlation function into  
\begin{eqnarray}
\Pi_{\mu\nu}=\frac{\langle 0 | J_{J/\psi,\mu}(0)| J/\psi(p',\varepsilon)\rangle \langle J/\psi(p',\varepsilon)|J^{tr,V;A}_{\nu}(0)|B_c(p)\rangle\langle B_c(p)|J_{B_c}(0)|0\rangle}{(p^2-m_{B_c}^2)(p'^2-m_{J/\psi}^2)}+\ldots,\label{Correl:1}
\end{eqnarray}
and
\begin{eqnarray}
\Pi_{\nu}=\frac{\langle 0 | J_{\eta_c}(0)| \eta_c(p')\rangle \langle \eta_c(p')|J^{tr,V;A}_{\nu}(0)|B_c(p)\rangle\langle B_c(p)|J_{B_c}(0)|0\rangle}{(p^2-m_{B_c}^2)(p'^2-m_{\eta_c}^2)}+\ldots,\label{Correl:2}
\end{eqnarray}
where the contributions of higher states and continuum are represented by $\ldots$. The matrix elements present in the above equations are parametrized in terms of masses, residues and the form factors as 
\begin{eqnarray}
\langle 0 | J_{B_c}|\eta_c(p)\rangle &=& -i \frac{m_{B_c}^2 f_{B_c}}{m_b+m_c},
\nonumber\\
\langle 0 | J_{J/\psi,\mu}|J/\psi(p',\varepsilon)\rangle &=&  \varepsilon_{\mu}m_{J/\psi}f_{J/\psi},\nonumber\\
\langle 0 | J_{\eta_c}|\eta_c(p')\rangle &=& -i \frac{m_{\eta_c}^2 f_{\eta_c}}{2m_c},
\nonumber\\
\langle J/\psi(p',\varepsilon)|J^{tr,V}_{\nu}(0)|B_c(p)\rangle & =& i\Big[f_0(q^2)(m_{B_{c}}+m_{J/\psi})\varepsilon^*_{\nu}-\frac{f_{+}(q^2)}{(m_{B_{c}}+m_{J/\psi})}(\varepsilon^* p)P_{\nu}-\frac{f_{-}(q^2)}{(m_{B_{c}}+m_{J/\psi})}(\varepsilon^* p)q_{\nu}\Big],\nonumber\\
\langle J/\psi(p',\varepsilon)|J^{tr,A}_{\nu}(0)|B_c(p)\rangle &=& \frac{f_{V}(q^2)}{(m_{B_{c}}+m_{J/\psi})}\epsilon_{\nu\delta\alpha\beta}\varepsilon^{*\delta} p^{\alpha}p'^{\beta} ,
\nonumber\\
\langle \eta_c(p')| J_{\nu}^{tr,V}(0)| B_c(p)\rangle &=& F_1(q^2) P_{\nu}+F_2(q^2) q_{\nu}.
\end{eqnarray}
Note that, for the transition including $\eta_c$ in the final state the axial vector part of the transition current does not contribute to the result due to the parity considerations. In the above expressions, $F_1(q^2)$, $F_2(q^2)$, $f_0(q^2)$, $f_{-}(q^2)$, $f_{+}(q^2)$ and $f_V(q^2)$ are the transition form factors; and $P_{\nu}= (p+p')_{\nu}$ and $q_{\nu} = (p-p')_{\nu}$. The use of the above matrix elements in Eqs.~(\ref{Correl:1}) and (\ref{Correl:2}) gives the final results of this side as
\begin{eqnarray}
\Pi_{\mu\nu}&=&\frac{f_{B_c} m_{B_c}^2}{m_b+m_c}\frac{f_{J/\psi}m_{J/\psi}}{(p^2-m_{B_c}^2)(p'^2-m_{J/\psi}^2)}\Big[f_0(q^2)g_{\mu\nu}(m_{B_{c}}+m_{J/\psi})-\frac{f_+(q^2)P_{\mu}p_{\nu}}{(m_{B_{c}}+m_{J/\psi})}-\frac{f_-(q^2)q_{\mu}p_{\nu}}{(m_{B_{c}}+m_{J/\psi})}\nonumber\\
&-&i\epsilon_{\alpha\beta\mu\nu}p^{\alpha}p'^{\beta}\frac{f_V(q^2)}{(m_{B_{c}}+m_{J/\psi})}\Big]+\ldots,\nonumber\\
\Pi_{\nu}&=&-\frac{1}{(p^2-m_{B_c}^2)(p'^2-m_{\eta_c}^2)}\frac{f_{\eta_c}m_{\eta_c}^2}{(2m_c)}\frac{f_{B_c} m_{B_c}^2}{m_b+m_c}\Big[F_1(q^2)P_{\nu}+F_2(q^2)q_{\nu}\Big]+\ldots.
\end{eqnarray} 
The form factors, $F_1(q^2)$, $F_2(q^2)$, $f_V(q^2)$, $f_{0}(q^2)$ and $f_{\pm}(q^2)$  will be extracted from the coefficients of the structures $P_{\nu}$, $q_{\nu}$, $\epsilon_{\alpha\beta\mu\nu}p^{\alpha}p'^{\beta}$, $g_{\mu\nu}$ and 
$\frac{1}{2}(p_{\mu}p_{\nu}\pm p'_{\mu}p_{\nu})$, respectively.

The second way to calculate the correlation function is done via application of the operator product expansion (OPE) in deep Euclidean region. In this side, the calculations are performed in terms of QCD degrees of freedom considering the interactions of the quarks and gluons in QCD vacuum. In this side, the explicit forms of the interpolating currents given in Eq.~(\ref{Currents}) are placed into the correlator. This is followed by the contraction of the quark fields using the Wick theorem. This application turns the correlators into
\begin{eqnarray}
\Pi_{\mu\nu}=i^2\int d^4x e^{-ipx} \int d^4y e^{ipy} Tr[\gamma_{\mu} S_c^{ij}(y)\gamma_{\nu}(1-\gamma_5)S_b^{jl}(-x)\gamma_5 S_c^{li}(x-y)],\label{Corjpsi}
\end{eqnarray}   
for the decay including $J/\psi$ and
\begin{eqnarray}
\Pi_{\nu}=i^2\int d^4x e^{-ipx} \int d^4y e^{ipy} Tr[\gamma_{5} S_c^{ij}(y)\gamma_{\nu}(1-\gamma_5)S_b^{jl}(-x)\gamma_5 S_c^{li}(x-y)],\label{Coretac}
\end{eqnarray}  
for the decay including $\eta_c$ in the final state. The $S_Q^{ij}$ in these results represents the heavy $c$ or $b$ quark propagator. Its explicit expression is given as~\cite{Ek4}
\begin{eqnarray}
S_{Q{ij}}(x)&=&\frac{i}{(2\pi)^4}\int d^4k e^{-ik \cdot x} \left\{
\frac{\delta_{ij}}{\!\not\!{k}-m_Q}
-\frac{g_sG^{\alpha\beta}_{ij}}{4}\frac{\sigma_{\alpha\beta}(\!\not\!{k}+m_Q)+
(\!\not\!{k}+m_Q)\sigma_{\alpha\beta}}{(k^2-m_Q^2)^2}\right.\nonumber\\
&&\left.+\frac{\pi^2}{3} \langle \frac{\alpha_sGG}{\pi}\rangle
\delta_{ij}m_Q \frac{k^2+m_Q\!\not\!{k}}{(k^2-m_Q^2)^4}
+\cdots\right\}.
\end{eqnarray}
 Although the nonperturbative parts containing gluon condensates provide very small contributions, we include these nonperturbative effects beside the perturbative ones.  The calculation of the perturbative part of the correlator is done using the Cutkosky rules~\cite{Cutkosky1960} in which the propagators having the forms $\frac{1}{p^2-m^2}$ are replaced by Dirac delta functions, $-2\pi\delta(p^2-m^2)$, implying that all quarks are real. After placing the propagators and performing the present integrals, the QCD sides emerge in terms of different Lorentz structures as 
 \begin{eqnarray}\label{QCDsidestructures:1}
\Pi^{QCD}_{\mu\nu}
&=&\Big(\Pi^{pert}_V(q^2)+\Pi^{non-pert}_V(q^2)\Big)\epsilon_{\mu\nu\alpha\beta}p'^{\alpha}p^{\beta}+
\Big(\Pi^{pert}_0(q^2)+\Pi^{non-pert}_0(q^2)\Big)g_{\mu\nu}\nonumber \\
&+&
\frac{1}{2}\Big(\Pi^{pert}_+(q^2)+\Pi^{non-pert}_+(q^2)\Big)(p_{\mu}p_{\nu}+p'_{\mu}p_{\nu})+\frac{1}{2}
\Big(\Pi^{pert}_{-}(q^2)+\Pi^{non-pert}_{-}(q^2)\Big)(p_{\mu}p_{\nu}-p'_{\mu}p_{\nu})
\nonumber \\
&+&\mbox{other\,\,\, structures},
\end{eqnarray}
 \begin{eqnarray}\label{QCDsidestructures:2}
\Pi^{QCD}_{\nu}
&=&\Big(\Pi^{pert}_1(q^2)+\Pi^{non-pert}_1(q^2)\Big)P_{\nu}+
\Big(\Pi^{pert}_2(q^2)+\Pi^{non-pert}_2(q^2)\Big)q_{\nu}.
\end{eqnarray}
The imaginary parts of the results obtained for perturbative parts, that is $\frac{1}{\pi}Im[\Pi_i^{pert}]$ where $i=V,0,+,-$ for transition of $B_c$ to $J/\psi$ and $i=1,2$ for transition of $B_c$ to $\eta_c$, give the spectral densities that are used in the following dispersion relation
\begin{eqnarray}\label{QCDside1}
\Pi^{pert}_i(q^2)=-\frac{1}{(2\pi)^2}\int^{}_{}ds\int^{}_{}ds'
\frac{\rho_i(s,s',q^2)}{(s-p^2)(s'-p'^2)}.
\end{eqnarray}
The results obtained for the spectral densities for the perturbative parts from the coefficients of the above structures are as follows:
\begin{eqnarray}
\rho_0(s,s')&=&6
 \Big[m_c q^2 - m_c s - m_c s' - 
    4 (m_b - m_c) C(q^2) - (m_b - m_c) (q^2 - s - s')A(q^2) + 
    2 (m_b-m_c) s' B(q^2) \Big] I_0(s,s',q^2),\nonumber\\
\rho_{+}(s,s')&=& 6 \Big[m_c - (m_b - 3 m_c) A(q^2) - 2 (m_b - m_c) D(q^2) + 2 m_c B(q^2) - 
    2 (m_b-m_c) E(q^2))\Big] I_0(s,s',q^2),\nonumber\\
 \rho_{-}(s,s')&=& 6\big[-m_c + (m_b + m_c) A(q^2) - 2 (m_b - m_c) D(q^2) - 2 m_c B(q^2) + 
   2( m_b-m_c) E(q^2) \Big] I_0(s,s',q^2),\nonumber\\
\rho_{V}(s,s')&=&12 \Big[m_c + (m_c-m_b) A(q^2)\Big] I_0(s,s',q^2),\nonumber\\
 \rho_1(s,s')&=&6  \Big[m_c (m_c - m_b) + s A(q^2) + s' B(q^2)\Big] I_0(s,s',q^2),\nonumber\\
 \rho_2(s,s')&=&6 \Big[(m_c - m_b) m_c + s A(q^2) - s' B(q^2)\Big] I_0(s,s',q^2), 
\end{eqnarray}
where the functions in spectral densities are defined as
\begin{eqnarray}
\lambda(a,b,c)&=&a^2+b^2+c^2-2ab-2ac-2bc,\nonumber\\
I_0(s,s',q^2)&=&\frac{1}{4\lambda^{\frac{1}{2}}(s,s',q^2)},\nonumber\\
A(q^2)&=& \frac{1}{\lambda(s,s',q^2)}(q^2-2m_b^2 + 2m_c^2 + s - s')s',\nonumber\\
B(q^2)&=& \frac{1}{\lambda(s,s',q^2)}\Big(2s s' + (m_b^2 - m_c^2 - s)(s + s'-q^2 )\Big),\nonumber\\
C(q^2)&=& \frac{}{2\lambda(s,s',q^2)}\Big(m_c^4s' + [m_b^4 + q^2s - m_b^2(q^2 + s - s')]s' + m_c^2[q^4 + s^2 - 2m_b^2 s' - s s' - q^2(2 s + s')]\Big),\nonumber\\
D(q^2)&=& \frac{1}{\lambda ^2(s,s',q^2)}\Big(s' \{6 m_c^4 s' +2 m_c^2 [(q^2 - s)^2 + (-6 m_b^2 + q^2 + s) s' - 2 s'^2] + s' [6 m_b^4 + q^4 + 4 q^2 s + s^2 \nonumber\\
&-& 6 m_b^2 (q^2 + s - s') - 2 (q^2 + s) s' + s'^2]\}\Big),\nonumber\\
E(q^2)&=&\frac{1}{\lambda^2(s,s',q^2)}\Big(m_c^2 (q^2 - s)^3 + (q^2 - s) [3 m_b^4 + 3 m_c^4 - m_c^2 (q^2 - 3 s) +s (2 q^2 + s) - 2 m_b^2 (3 m_c^2 + q^2 + 2 s)] s' \nonumber\\
&-& [3 m_b^4 +3 m_c^4 + m_c^2 (q^2 - 3 s) + (q^2 - 2 s) s +2 m_b^2 (-3 m_c^2 - 2 q^2 + s)] s'^2 + (-2 m_b^2 + m_c^2 - s) s'^3\Big).
\end{eqnarray}
The three $\delta$ functions in the calculations determine the integration regions for the perturbative calculations. With the condition that the argument of the $\delta$ functions vanish simultaneously the following non-equality is obtained:
\begin{eqnarray}
-1\leq f(s,s')= \frac{2s s'+(s+s'-q^2)(m_b^2-s-m_c^2)}{\lambda^{1/2}(m_b^2,s,m_c^2)\lambda^{1/2}(s,s',q^2)}\leq +1,
\end{eqnarray}  
which describes the physical region in the $s$ and $s'$ plane. 
As for the calculations of the nonperturbative contributions, we apply the Schwinger representation of the Euclidean propagator together with Gaussian integrals to calculate the integrals present in these parts. The results for these contributions are very lengthy and therefore we do not give their explicit forms here.

After getting the results for phenomenological and QCD sides, the coefficients of the same Lorentz structures, selected from both sides, are matched to attain the sum rules of the form factors which are as follows
\begin{eqnarray}
f_i(q^2)&=&\xi\frac{(m_b+m_c)e^{\frac{m_{B_c}^2}{M^2}} e^{\frac{ m_{J/\psi}^2}{M'^2}}\Delta}{
    f_{B_c} f_{J/\psi} m_{B_c}^2 m_{J/\psi }}\Bigg\{
    -\frac{1}{(2\pi)^2}\int_{(m_b+m_c)^2}^{s_0}ds\int_{4m_c^2}^{s_0'}ds'\rho_i({s,s',q^2})\theta[1-f^2(s,s')]e^{\frac{-s}{M^2}}e^{\frac{-s'}{M'^2}}\nonumber\\
    &+&\hat{\textbf{B}}\Pi_i^{\mbox{non-pert.}}\Bigg\},\label{QCDsumrule1}
\end{eqnarray}
  for the $B_c\rightarrow J/\psi l \nu$ decay, and 
  \begin{eqnarray}
F_{1,2}(q^2)&=&-\frac{(m_b+m_c)2m_c e^{\frac{m_{B_c}^2}{M^2}} e^{\frac{ m_{\eta_c}^2}{M'^2}}}{
    f_{B_c} f_{\eta_c} m_{B_c}^2 m_{\eta_c }^2}
   \Bigg\{ -\frac{1}{(2\pi)^2}\int_{(m_b+m_c)^2}^{s_0}ds\int_{4m_c^2}^{s_0'}ds'\rho_{1,2}({s,s',q^2})\theta[1-f^2(s,s')]e^{\frac{-s}{M^2}}e^{\frac{-s'}{M'^2}}\nonumber\\
    &+&\hat{\textbf{B}}\Pi_{1,2}^{\mbox{non-pert.}}\Bigg\},\label{QCDsumrule2}
\end{eqnarray}
for the $B_c\rightarrow \eta_c l \nu$ decay. The sub-index $i$ in the form factors of $B_c\rightarrow J/\psi l \nu$ decay is $i=0,+,-,V$ as we previously mentioned.  In this channel,  $\Delta=\frac{1}{m_{B_c}+m_{\j/\psi}}$ for $i=0$ and $\Delta=m_{B_c}+m_{\j/\psi}$ for $i=+,-,V$. Here, $\xi=+1$ for $i=0$ and $\xi=-1$ for $i=+,-,V$. The QCD sum rule equations contain also the contributions from higher states and continuum. To subtract these unwanted contributions we apply quark hadron duality assumption and for their further suppression double Borel transformation is used with respect to the variables $p^2$ and $p'^2$. The results given in ~Eqs.(\ref{QCDsumrule1}) and (\ref{QCDsumrule2}) are those obtained after the quark hadron duality assumption and double Borel transformation.   

\section{Numerical Analyses}

The results obtained from the QCD sum rules calculations in the previous section are numerically analyzed in this section with the usage of the input parameters given as $m_b=4.18^{+0.04}_{-0.03}$~GeV, $m_c=1.275^{+0.025}_{-0.035}$~GeV, $m_{B_c} = 6274.9\pm 0.8$~MeV, $m_{J/\psi} = 3096.900\pm 0.006$~MeV, $m_{\eta_c}=2983.9 \pm 0.5$~MeV, $\tau_{B_c}=(0.507\pm 0.009)\times 10^{-12}$~s~\cite{Tanabashi2018}, $f_{B_c} = 400\pm 45$~MeV~\cite{Kiselev:2000jc}, $f_{J/\psi}=411 \pm 7$~MeV~\cite{Dudek:2006ej}, and $f_{\eta_c} = 300 \pm 50$~MeV~\cite{Deshpande:1994mk}. 

\begin{figure}%
\centering
\subfigure[][]{%
\label{fig:ex3-a}%
\includegraphics[height=2in]{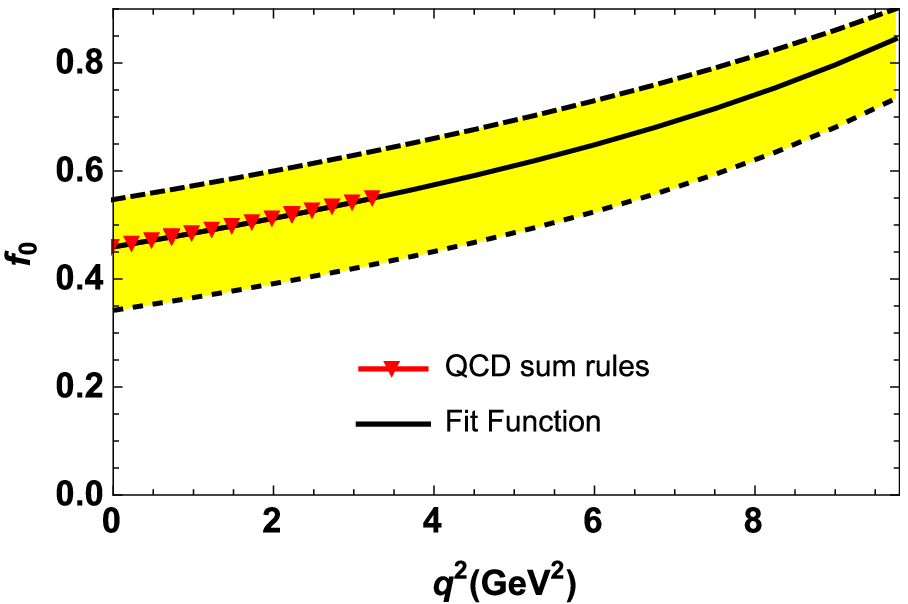}}%
\hspace{8pt}%
\subfigure[][]{%
\label{fig:ex3-b}%
\includegraphics[height=2in]{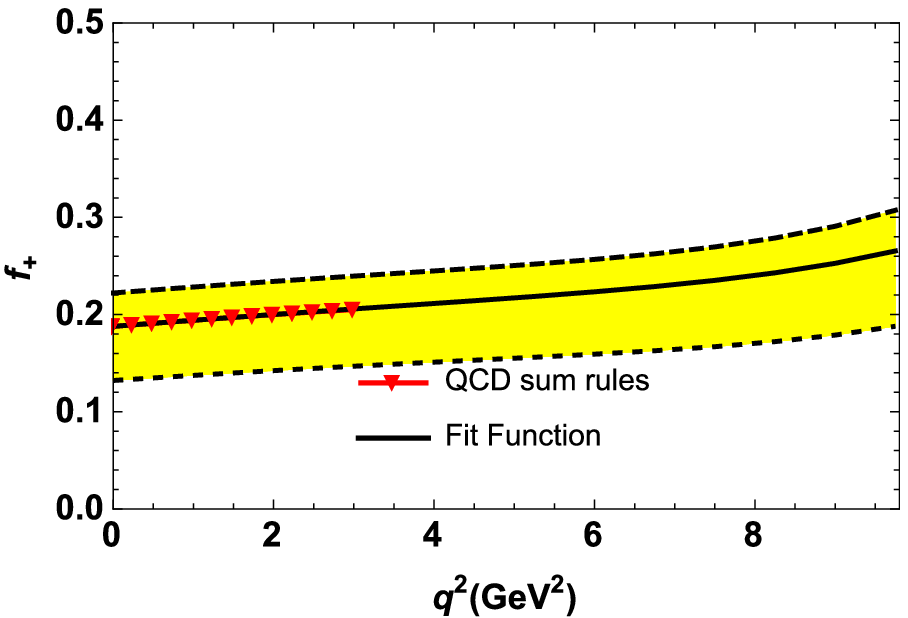}} \\
\subfigure[][]{%
\label{fig:ex3-c}%
\includegraphics[height=2in]{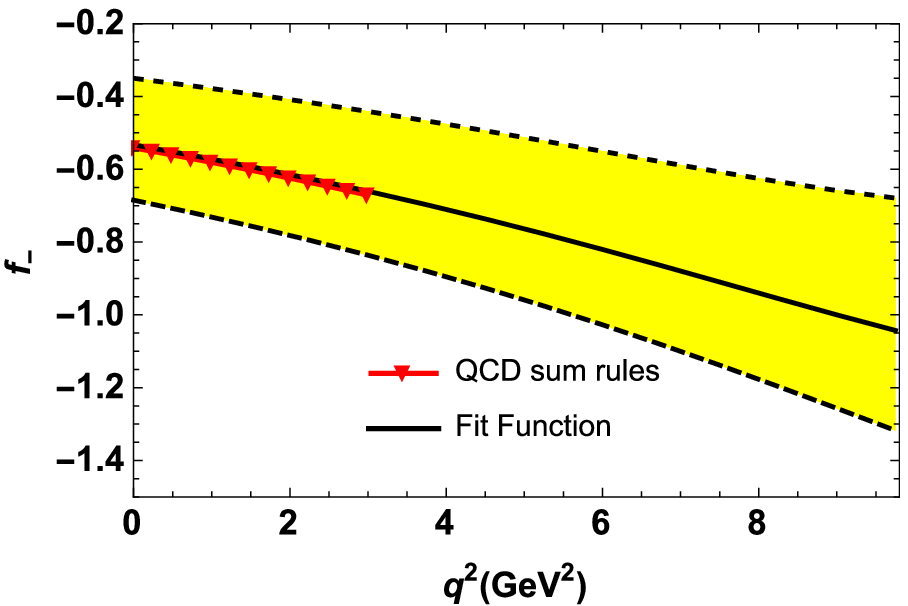}}%
\hspace{8pt}%
\subfigure[][]{%
\label{fig:ex3-d}%
\includegraphics[height=2in]{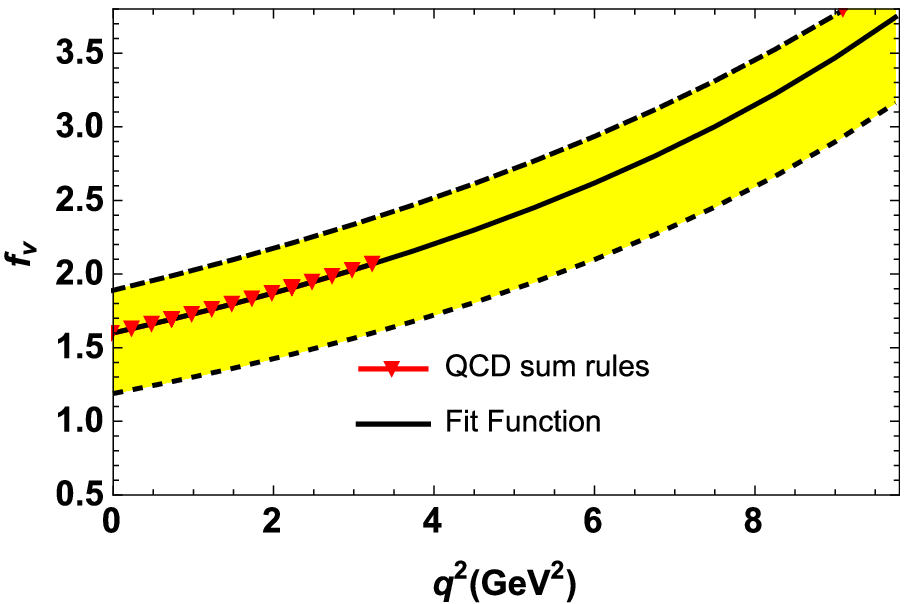}}%
\caption[A set of four subfigures.]{The variations of the form factors as functions of $q^2$ for the $B_c\rightarrow J/\psi l \nu$ at central values of the input parameters $M^2$, $M'^2$, $s_0$ and $s_0'$. The red triangles present the results obtained from QCD sum rule calculations. The black solid lines are the results of fit functions obtained using central values of the input parameters. The yellow bands indicate the errors arising from the variations of the input parameters.:
\subref{fig:ex3-a} For $f_{0}(q^2)$;
\subref{fig:ex3-b} For $f_{+}(q^2)$;
\subref{fig:ex3-c} For $f_{-}(q^2)$; and,
\subref{fig:ex3-d} For $f_{V}(q^2)$.}%
\label{fig:Jpsi}%
\end{figure}

\begin{figure}%
\centering
\subfigure[][]{%
\label{fig:ex3-a}%
\includegraphics[height=2in]{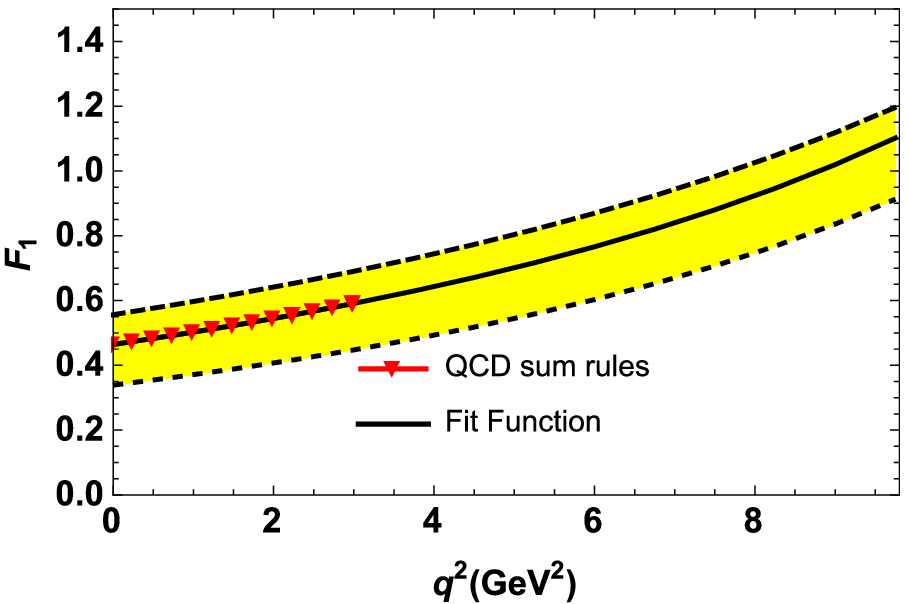}}%
\hspace{8pt}%
\subfigure[][]{%
\label{fig:ex3-b}%
\includegraphics[height=2in]{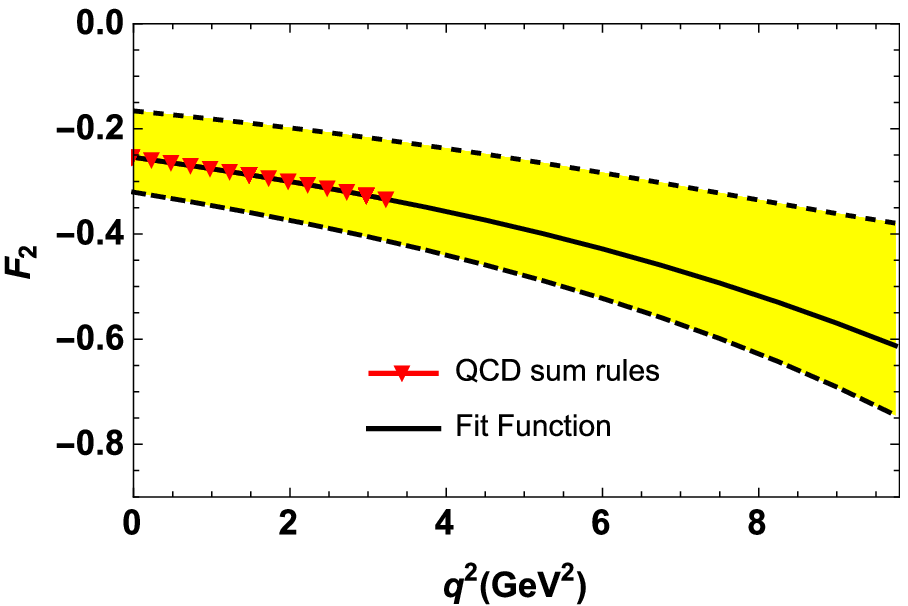}}%
\caption[A set of four subfigures.]{The variations of the form factors as functions of $q^2$ for the $B_c\rightarrow \eta_c l \nu$ at central values of the input parameters $M^2$, $M'^2$, $s_0$ and $s_0'$. The red triangles present the results obtained from QCD sum rule calculations. The black solid lines are the results of fit functions obtained using central values of the input parameters. The yellow bands indicate the errors arising from the variations of the input parameters.:
\subref{fig:ex3-a} For $F_{1}(q^2)$; and,
\subref{fig:ex3-b} For $F_{2}(q^2)$.}%
\label{fig:Etac}%
\end{figure}

These are not the only parameters needed in the calculations. There are four auxiliary parameters which are Borel parameters, $M^2$, $M'^2$, and the threshold parameters, $s_0$ and $s_0'$. Demanding weak dependency of the results on these parameters, their working intervals are fixed. The exact upper and lower bounds of them are set considering the criteria of the QCD sum rules. These criteria include the pole dominance as well as convergence of the QPE, that is, the perturbative contribution prevails over the nonperturbative ones and the higher  the dimension of nonperturbative operator the lower is its contribution. By imposing the condition of OPE convergence, we achieve the  lower limit of the Borel parameters. To attain the upper limit for the Borel parameters the criterion is the pole dominance. Considering that the pole contribution consists at least $50\%$ of the total result, we adjust the upper limits of the Borel parameters. Hence,   we get
\begin{eqnarray}
&6~\mbox{GeV}^2\leq M^2\leq 10~\mbox{GeV}^2, &\nonumber\\
\end{eqnarray}
and
\begin{eqnarray}
&4~\mbox{GeV}^2 \leq M'^2\leq 6~\mbox{GeV}^2 &.
\end{eqnarray}
The threshold parameters have relations with the energies of the first excited states in the initial and final channels, and therefore are chosen as
\begin{eqnarray}
&43~\mbox{GeV}^2 \leq s_0 \leq 48~\mbox{GeV}^2, &
\end{eqnarray}
and
\begin{eqnarray}
& 11~\mbox{GeV}^2 \leq s'_0 \leq 15~\mbox{GeV}^2 &.
\end{eqnarray}
To attain the decay widths of the considered decays it is necessary to have the form factors describing these decays as functions of the $q^2$ in the whole physical region, that is, $m_l^2 \leq q^2 \leq (m_{B_c}-m_{J/\psi(\eta_c)})^2$. However, in our analyses we encounter that the form factors truncate at some $q^2$ values. Therefore, to extend them to the whole physical region it is required to use suitable fit functions having same behaviors with our QCD sum rule results in the regions that our results are valid. The fit functions used in these calculations have the following form:
\begin{eqnarray}
f_i(q^2)=\frac{f_i(0)}{1+a_1\hat{q}+ a_2 \hat{q}^2+ a_3 \hat{q}^3 +a_4 \hat{q}^4},
\end{eqnarray}
where $\hat{q}$ in the above fit function is expressed as $\hat{q}=\frac{q^2}{m_{B_c}^2}$. Our analyses lead to the values of the parameters of the fit functions which are given in the Table~\ref{paramtersoffits}. The $+,-$ values contained in the results are indicating the upper and lower bounds for the values of the fit parameters obtained in the analyses.  
\begin{table}[]
\begin{tabular}{|c|c|c|c|c|c|}
\hline
      & $f_i(0)$            & $a_1$      & $a_2$     & $a_3$       & $a_4$  \\ \hline\hline
$f_0$ & $0.46^{+0.12}_{-0.09}$    & $-2.11^{+0.32}_{-0.50}$ &$1.52^{+0.46}_{-1.02}$ &$-1.15^{+0.49}_{-1.32}$  & $-2.43^{+0.75}_{-0.01}$  \\ \hline
$f_+$ & $0.19^{+0.06}_{-0.03}$    & $-1.34^{+0.23}_{-0.27}$ & $2.91^{+1.13}_{-0.61}$& $-1.51^{+0.46}_{-0.01}$  & $-30.48^{+12.43}_{-12.03}$ \\ \hline
$f_-$ & $-0.57^{+0.19}_{-0.14}$   & $-2.78^{+0.22}_{-0.26}$ & $3.25^{+1.20}_{-0.71}$ & $-1.77^{+0.71}_{-1.18}$  & $7.09^{+3.39}_{-3.22}$ \\ \hline
$f_V$ &   $1.60^{+0.29}_{-0.41}$  & $-3.03^{+0.32}_{-0.49}$ & $3.48^{+1.50}_{-0.79}$ & $-2.49^{+0.93}_{-2.14} $ & $0.29^{+2.08}_{-0.01}$ \\ \hline\hline
$F_1$ &   $0.46^{+0.09}_{-0.13}$  & $-3.07^{+0.32}_{-0.48}$ & $3.60^{+1.53}_{-0.82}$ & $-2.63^{+0.99}_{-2.20}$  & $0.40^{+2.18}_{-0.32}$ \\ \hline
$F_2$ &   $-0.25^{+0.09}_{-0.07}$ & $-3.20^{+0.23}_{-0.25}$ & $3.82^{+1.34}_{-0.77}$ & $-2.66^{+0.93}_{-1.43}$  & $3.69^{+6.20}_{-2.62}$ \\ \hline
\end{tabular}
\caption{The parameters of the fit functions obtained for $B_c\rightarrow J/\psi l \nu$ and $B_c\rightarrow \eta_c l \nu$ decays at central values of the Borel and threshold parameters.}
\label{paramtersoffits}
\end{table}
The fit functions and results of the QCD sum rule calculations are depicted as functions of $q^2$ in Figs.~\ref{fig:Jpsi} and \ref{fig:Etac}, to show the consistency of them in the working regions of the QCD sum rule analyses. It can be seen from these figures that the chosen fit functions have good overlap with the results of the QCD sum rule calculations in viable regions and therefore can be used to enlarge the results to the whole physical region.  In these figures, the lines drawn with red triangles show the results obtained from the QCD sum rule calculations for the form factors. The solid black lines indicate the fit functions obtained for the form factors using the central values of the auxiliary parameters. And, the uncertainties present in the predictions because of variations of the input parameters are pointed out by the yellow bands.    
The obtained fit functions are used in the next step to calculate the corresponding decay widths. For the decay width of the $B_c\rightarrow J\psi l \nu $ we use the decay width formula given in Ref.~\cite{Ivanov:2000aj}, and as for that of the 
$B_c\rightarrow \eta_c l \nu $ we adopt the formula given in Ref.~\cite{Issadykov:2018myx}. 
\begin{table}[]
\begin{tabular}{|c|c|c|c|c|c|c|c|c|c|c|}
\hline 
Mode   & This Work   &\cite{Issadykov:2018myx}&\cite{Leljak:2019eyw}&\cite{Huang:2007kb}&\cite{Kiselev:2002vz}&\cite{Ivanov:2006ni}&\cite{Wen-Fei:2013uea}&\cite{Wang:2008xt} & \cite{Hu:2019qcn} \\ \hline  \hline
$BR(B_c\rightarrow J/\psi \mu \nu)$ & $1.93^{+0.50}_{-0.60}$ &$1.67\pm 0.33$&$2.24^{+0.57}_{-0.49}$ & ~$2.37$~ &~$1.9$~&~$2.07$~&~$1.003^{+0.133}_{-0.118}$~&~$1.49^{+0.01+0.15+0.23}_{-0.03-0.14-0.23}$~& $ 0.998^{+0.065}_{-0.018}$ \\ \hline
$BR(B_c\rightarrow J/\psi \tau \nu)$&$0.49^{+0.10}_{-0.14} $&$ 0.40\pm 0.08$& $0.53^{+0.16}_{-0.14}$ & $0.65$ &$0.48$&$0.49$&$0.292^{+0.040}_{-0.034}$&$0.370^{+0.002+0.042+0.056}_{-0.005-0.038-0.056}$ & $ 0.230^{+0.060}_{-0.038}$ \\ \hline
$BR(B_c\rightarrow \eta_c \mu \nu)$ &$0.56^{+0.19}_{-0.23} $&$ 0.95\pm0.19$&$0.82^{+012}_{-0.11}$&$1.64$ &$0.75$&$0.81$&$0.441^{+0.122}_{-0.109}$&$0.67^{+0.04+0.04+0.10}_{-0.07-0.04-0.10}$ & $ 0.720^{+0.180}_{-0.140}$  \\ \hline
$BR(B_c\rightarrow \eta_c \tau \nu)$& $0.21^{+0.04}_{-0.06}$ &$0.24\pm0.05$&$0.26^{+0.06}_{-0.05}$ &$0.49$&$0.23$&$0.22$&$0.137^{+0.037}_{-0.034}$&$0.190^{+0.005+0.014+0.029}_{-0.012-0.013-0.029}$& $ 0.216^{+0.030}_{-0.025}$ \\ \hline
$R(J/\psi)$ & $0.25^{+0.01}_{-0.01}$& $0.24\pm0.05$  & $0.23\pm 0.01$& $0.27$ &$0.25$&$0.24$&$0.29$&$0.25$ & $ 0.230^{+0.041}_{-0.035}$ \\ \hline
$R(\eta_c)$ &$ 0.36^{+0.05}_{-0.03}$&$0.26\pm0.05$ &$0.32\pm 0.02$ &$0.30$&$0.31$&$0.27$&  $0.31$&$0.28$& $ 0.300^{+0.033}_{-0.031}$  \\ \hline \hline
\end{tabular}
\begin{tabular}{|c|c|c|c|c|c|c|c|c|c|c|}
Mode  & \cite{Huang:2018nnq} & \cite{Colangelo:1999zn}   &\cite{Murphy:2018sqg}&\cite{Berns:2018vpl}&\cite{Colangelo:1}&\cite{Alok:2017qsi}\\ \hline  \hline
$BR(B_c\rightarrow J/\psi \mu \nu)$&- & $1.5(3.3)$ &-&- & ~$0.84$~&- \\ \hline
$BR(B_c\rightarrow J/\psi \tau \nu)$&- &-&-& - & -&- \\ \hline
$BR(B_c\rightarrow \eta_c \mu \nu)$&- &$0.15(0.5)$& -&-& 0.17&- \\ \hline
$BR(B_c\rightarrow \eta_c \tau \nu)$&- &-& - &-&- &- \\ \hline
$R(J/\psi)$   &$0.248(6)(0)$  & -& $ 0.26 \pm 0.02$  & -& - &$0.289\pm 0.007$\\ \hline
$R(\eta_c)$ &$0.281^{+0.034}_{-0.030}(0)$  &-&$0.31^{+0.04}_{-0.02}$ &$0.29(5)$ &- &- \\ \hline 
\end{tabular}
\caption{The branching fractions in \% for $B_c\rightarrow J/\psi l \nu$ and $B_c\rightarrow \eta_c l \nu$, as well as  $R(J/\psi)$ and $R(\eta_c)$.}
\label{tab:BR}
\end{table}
Table~\ref{tab:BR} presents the results of the $BR$s that we obtain in this work together with the results obtained in some other studies. In our results, we present the errors of our calculations in which we consider the effects of variations inherited by the variations of the form factors. In this table we also present the ratios of the $BR$s , i.e., $R(J/\psi)$ and $R(\eta_c)$ and compare the results with other theoretical predictions.

\section{Discussion and Conclusions}

The present work includes analyses on the semileptonic decays $B_c\rightarrow J/\psi l \nu$ and $B_c\rightarrow \eta_c l \nu$. Motivated by the recent observation of the LHCb~~\cite{Aaij:2017tyk} on $R(J/\psi)$, indicating serious deviations of the experimental data from the existing SM predictions and the possibility of new physics effects, we first calculated the form factors entering the amplitude for the hadronic matrix elements of the $B_c\rightarrow J/\psi l \nu$ transition. We applied the three-point QCD sum rule approach to find the fit functions of the form factors defining  the tree-level transition of $B_c\rightarrow J/\psi l \nu$ in terms of $q^2$ in whole physical region. We used these functions to estimate the $BR$s of the $B_c\rightarrow J/\psi l \nu$ in $\tau$ and $\mu$ channels. The obtained $BR$s, $BR(B_c\rightarrow J/\psi \mu \nu) = 1.93^{+0.50}_{-0.60}$ and $BR(B_c\rightarrow J/\psi \tau \nu)=0.49^{+0.10}_{-0.14} $, agree with most of the present theoretical findings, as is seen from Table~ \ref{tab:BR}, within the error intervals of the predictions. Although our result of $BR$ for $B_c\rightarrow J/\psi \mu \nu$ is slightly larger than those of Refs.~\cite{Wen-Fei:2013uea,Hu:2019qcn,Colangelo:1}, it is consistent with the results obtained in Refs.~\cite{Issadykov:2018myx,Leljak:2019eyw,Huang:2007kb,Kiselev:2002vz,Ivanov:2006ni,Wang:2008xt,Colangelo:1999zn} within the uncertainties. When the $B_c\rightarrow J/\psi \tau \nu$ is considered, our result shows small differences with the predictions of Refs.~\cite{Huang:2007kb,Wen-Fei:2013uea,Hu:2019qcn}, while it is consistent with the results of the Refs.~\cite{Issadykov:2018myx,Leljak:2019eyw,Kiselev:2002vz,Ivanov:2006ni,Wang:2008xt} within the errors. We obtained the corresponding $R(J/\psi)$ as $R(J/\psi)=\frac{BR(B_c\rightarrow J/\psi \tau \nu)}{BR(B_c\rightarrow J/\psi \mu \nu)}=0.25^{+0.01}_{-0.01}$, which shows small differences with the predictions of the  Refs.~\cite{Wen-Fei:2013uea,Huang:2007kb}.   $R(J/\psi)$ values obtained in Refs.~\cite{Wen-Fei:2013uea,Huang:2007kb} are slightly larger than our prediction. On the other hand, our result is in agreement with those of Refs.~\cite{Leljak:2019eyw,Kiselev:2002vz,Wang:2008xt,Issadykov:2018myx,Ivanov:2006ni,Hu:2019qcn,Huang:2018nnq,Murphy:2018sqg,Alok:2017qsi} within the errors. Our prediction for  $R(J/\psi)$ differs considerably with  the LHCb result, $R(J/\psi)=0.71(17)(18)$~\cite{Aaij:2017tyk}, indicating serious  LFUV.\\
   
We also considered the possibility of the future similar measurements for the $B_c\rightarrow \eta_c l \nu$ channel and calculated corresponding form factors. The results obtained for the form factors are used to obtain the related $BR$s and $R(\eta_c)$. Our predictions are as follows: $BR(B_c\rightarrow \eta_c \mu \nu) = 0.56^{+0.19}_{-0.23} $ and $BR(B_c\rightarrow \eta_c \tau \nu)= 0.21^{+0.04}_{-0.06}$ giving the ratio  $R(\eta_c)=\frac{BR(B_c\rightarrow \eta_c \tau \nu)}{BR(B_c\rightarrow \eta_c \mu \nu)}=0.36^{+0.05}_{-0.03}$. If we compare our $BR$ results with those of the references given in Table~\ref{tab:BR}, the result obtained for $B_c\rightarrow \eta_c \mu \nu$ is consistent with those of Refs.~\cite{Leljak:2019eyw,Kiselev:2002vz,Wen-Fei:2013uea,Wang:2008xt,Hu:2019qcn,Colangelo:1999zn} within the errors, however, it is considerably different than the results of Refs.~\cite{Issadykov:2018myx,Huang:2007kb,Ivanov:2006ni,Colangelo:1}. As for $B_c\rightarrow \eta_c \tau \nu$, a considerable difference is present between our result and  that of Ref.~\cite{Huang:2007kb}, while the Refs.~\cite{Issadykov:2018myx,Leljak:2019eyw,Kiselev:2002vz,Ivanov:2006ni,Wang:2008xt,Wen-Fei:2013uea,Hu:2019qcn}  have predictions, which are in consistency within the errors with our result. Our result on $R(\eta_c)$ is slightly different than those of Refs.~\cite{Huang:2007kb,Kiselev:2002vz,Ivanov:2006ni,Wen-Fei:2013uea,Wang:2008xt,Issadykov:2018myx,Huang:2018nnq}  while it is consistent with the predictions of~Refs.~\cite{Leljak:2019eyw,Hu:2019qcn,Murphy:2018sqg,Berns:2018vpl}, when their errors are considered.  Note that for those predictions that do not contain the uncertainties of the results, the central values have been considered in making the above conclusions. \\

Our results on $R(J/\psi)$ and $R(\eta_c)$ contain $ 4\% $ and $ (8-14)\% $ errors, respectively. Considering for instance the errors of the form factors  at $ q^2=0 $ in Table~\ref{paramtersoffits}, which are in the order of $ (16-33)\% $ and $ (20-36)\% $ respectively for $B_c\rightarrow J/\psi l \nu$ and $B_c\rightarrow \eta_c l \nu$ channels, we see considerable cancellations of the  theoretical uncertainties in the ratios. Similar cancellations at both channels are  occurred in the results of Refs. \cite{Issadykov:2018myx,Leljak:2019eyw,Hu:2019qcn,Huang:2018nnq,Murphy:2018sqg,Berns:2018vpl,Alok:2017qsi}, where the errors of the results were presented. We shall note that only the prediction of Ref. \cite{Cohen:2018dgz}   on $ R(J/\psi) $, $0.20\leqslant R(J/\psi)\leqslant 0.39$, represents a wide band, which was obtained   by constraining the form factors through a combination of dispersive relations, heavy-quark relations at zero-recoil, and the limited existing determinations from lattice QCD with different sources of uncertainties. Our prediction on $R(J/\psi)$ is more precise compared to that of  $R(\eta_c)$. This can be attributed to the fact that the values of the form factors presented in Table~\ref{paramtersoffits} are more uncertain in $B_c\rightarrow \eta_c l \nu$ channel compared to the $B_c\rightarrow J/\psi l \nu$ mode. The main reason behind this is that our knowledge on the parameters of $ \eta_c $ is poor compared to those of $ J/\psi $ channel. For instance the uncertainty in the value of the decay constant for $ \eta_c $,  $f_{\eta_c} = 300 \pm 50$~MeV, which enters as one of the main inputs to the expressions of the sum rules in Eq. (\ref{QCDsumrule2}), is very high compared to  that of $f_{J/\psi}=411 \pm 7$~MeV. This is the case regarding the experimental values of the masses for these two quarkonia as other inputs of form factors. The experimental value for the mass of $ \eta_c $ presented in the beginning of the previous section suffers from large uncertainty compared to the experimental value for the mass of $ J/\psi $ meson. As we also previously mentioned,  predictions on  $ R(D) $ and $ R(D^*) $ are more precise compared to those of $ R(J/\psi) $. This is because of the fact that our theoretical and experimental knowledge on the parameters of $ B $, $ D $ and $ D^* $ mesons, which are entered as inputs to the expressions of the form factors, are overall more precise compared to the parameters of the $ B_c $ and $ J/\psi $ mesons.

As it can be seen, our result on $R(J/\psi)$ supports the present tension between SM theory predictions and experiment which indicates that it is necessary to have more precise experimental data to account for this discrepancy. On the other hand, similar future experimental measurements on $R(\eta_c)$ may provide valuable information on the possible lepton universality violation in  $B_c\rightarrow \eta_c l \nu$ channel. The results of our study and the other theoretical predictions can be useful in this respect.


\end{document}